\def\etal{{\rm et al. }}
\def\mpc{{h^{-1} \rm Mpc}}
\def\kpc{{h^{-1} \rm kpc}}
\def\kms{{\rm km s^{-1}}}
\newcommand\aap{{\em A}\&{\em A}}

\newcommand\aj{{\em AJ}}
\newcommand\apj{{\em ApJ}}

\newcommand\apjs{{\em ApJS}}

\newcommand\mn{{\em MNRAS}}

\newcommand\nat{{\em Nature}}
\newcommand\nature{{\em Nature}}
\newcommand\pasp{{\em PASP}}
\newcommand\science{{\em Science}}

\documentclass[usenatbib]{mn2e}

\usepackage[dvips]{graphicx}
\begin{document}

\title{Dichotomy in host environments and signs of recycled AGN}

\author[Coldwell \etal]{Georgina V. Coldwell$^{1,2}$\thanks{E-mail:
georgina@mail.oac.uncor.edu (GVC)}, Diego G. Lambas$^{1}$, Ilona K. 
S\"ochting$^{3}$ and 
Sebasti\'an Gurovich$^{1}$\\
$^{1}$ IATE, Observatorio Astron\'omico, Universidad Nacional de C\'ordoba, 
Laprida 854, 5000, C\'ordoba, Argentina\\
$^{2}$ Departamento de F\'isica, Universidad de La Serena, Benavente 980, La Serena, Chile\\
$^{3}$ University of Oxford, Astrophysics, Denys Wilkinson Building, Keble 
Road, Oxford OX1 3RH, UK}

\date{\today}

\pagerange{\pageref{firstpage}--\pageref{lastpage}}

\maketitle

\label{firstpage}

\begin{abstract}
We analyse the relation between AGN host properties and large scale
environment for a representative red and blue AGN host galaxy sample
selected from the DR4 SDSS. A comparison is made with two carefully
constructed control samples of non-active galaxies, covering the same
redshift range and color baseline.  The cross-correlation functions show
that the density distribution of neighbours is almost identical for
blue galaxies, either active, or non-active. Although active red
galaxies inhabit environments less dense compared to non-active red
galaxies, both reside in environments considerably denser
than those of blue hosts. Moreover, the radial density profile of AGN,
relative to galaxy group centres is less concentrated than
galaxies. This is particularly evident when comparing red AGN and
non-active galaxies.

The properties of the neighbouring galaxies of blue and red AGN and
non active galaxies reflect this effect. While the neighbourhood of
the blue samples is indistinguishable, the red AGN environs show an
excess of blue-star forming galaxies with respect to their non-active
counterpart.  On the other hand, the active and non-active blue
systems have similar environments but markedly different
morphological distributions, showing an excess of blue early-type AGN,
which are argued to be late stage mergers. This comparison reveals that
the observable differences between active red and blue host galaxy
properties including star formation history and AGN activity depends
on the environment within which the galaxies form and evolve.
\end{abstract}


\section{Introduction}

The standard model of the Active Galactic Nuclei (AGN) phenomenon
includes the accretion of cold gas by massive black holes that lie at
the centres of galaxies \citep{LB69}. Although there is considerable
evidence to suggest that most massive galaxies harbour a central black
hole \citep{trem02,MH03}, not all massive galaxies in fact are
active. The search for the underlying reason on the observed occurrence
or absence of AGN in massive galaxy hosts has been intense and although
some progress has been made, largely this question remains
unanswered. Many of the details of the gas cooling and transportation
processes that trigger and produce AGN activity remain open to debate,
yet several processes that initialize and facilitate the gas transport
have been proposed.

\cite{TT72} suggest that collision driven disruption and gas dissipation may 
feed the nuclear activity of galaxies.  It is likely that tidal torques 
generated 
during galaxy interactions also play some role in providing gas inflows that 
end up feeding central black holes
\citep{sanders88} that power AGN. Nevertheless, recent studies 
\citep{coldwell06,Li06,alonso07} 
have found no statistical evidence that major galaxy interactions are
an efficient mechanism for the triggering of nuclear
activity. However, the effect that minor
mergers \citep{roos81,roos85,gaskell85,HM95} or galaxy
harassment \citep{LKM98} processes have on nuclear activity
remains yet to be quantified. Some other triggering mechanisms have been
suggested, for example, internal instabilities in the disks of barred
galaxies that cause gas transfer \citep{SBF90} as well as other gas
dynamical processes implied by the presence of multiple super-massive
black holes found in the centres of some active host
galaxies \citep{BBR80}.  One must also consider that it is plausible
that at any given time or for any given AGN type, more than one mechanism
plays a role. The relative contributions of these processes probably
evolve with redshift.

The environments of galaxies play a significant role in
their evolution and so understanding the relation of nuclear activity
with host galaxy environment provides an alternative angle to this
problem. So bringing all these possibilities together, the question
that has yet to be answered is if nuclear activity is a temporary
stage in the evolution, common to all galaxies of a certain type or if
it is induced by events outside the host galaxy (nature versus
nurture).  In the nature case, the environment of AGN hosts would be a
simple reflection of the environment typical to non-active galaxies of
similar type. Therefore, the environment of AGN host galaxies adds a much 
needed
constraint in understanding the AGN phenomenology. 

Considerable effort has gone into such an undertaking and some
progress has been made. \cite{Li06} compare the clustering of galaxies
centred about narrow-line galaxies (Type II AGN) to a matched
non-active centred host sample environ. They find no significant
difference in the clustering amplitude at a scale larger than $1 \mpc$
and only a weak signal of higher galaxy density in the AGN environs at
scales smaller than $1 \mpc$.
\cite{SRR06} study the density of AGN environs, star-forming galaxies 
and normal galaxies and find no evidence of a relation between large-scale 
environment and AGN
activity.  Several other studies also indicate that AGN and quasars
reside in environments of similar galaxy density as those of non AGN
galaxies \citep{SBM95,coldwell06}.
Nevertheless, the physical properties of the neighbouring galaxies
appear to be significantly different.

Interesting parallels are observed for low redshift quasars. Even
though the distribution of quasars follow the same general large-scale
structure traced by galaxy clusters, they are less concentrated around
galaxy cluster centres but instead, more likely to be found in the
periphery of clusters and between possibly merging, galaxy
clusters \citep{ilona02,ilona04}. It is precisely in these
environments that disks are more prevalent because dynamical effects
that tend to remove baryons and destroy disk formation (eg: ram
pressure stripping, tidal forces, galaxy harassment) that are more
pronounced in denser environments are less significant. Hence,
galaxies in both quasar and AGN environs are particularly blue, disky
and star-forming compared to those in the vicinity of typical galaxies
up to scales of $\sim 1 \mpc$ \citep{coldwell03,coldwell06}.

Could overdensities observed strictly locally
around quasars \citep{serber06} and AGN \citep{koulo06} within $\sim
0.1 \mpc$ cause AGN activity through gas dynamical galaxy interactions?
Interestingly, evidence to the affirmative have been presented by the
above authors, who find that at larger scales ($\sim 1 \mpc$), active
objects inhabit regions with densities similar to that found for $L^*$
galaxies, consistent with the results of \cite{coldwell06}.

\cite{biviano1997} suggest that the
lower AGN fraction in galaxy clusters originally proposed
by \cite{DTS85} is due to differences in the morphological mixture of
galaxies in cluster and field environments. \cite{miller03} on
the other hand find little evidence of environmental dependence on AGN
fraction by analyzing SDSS galaxies across a
wide range of environments. In spite of this, \cite{kauff04} also
analyze SDSS galaxies yet they find that AGN under-populate regions
of high galaxy density.

These seemingly contradictory findings indicate that the study of AGN
environs is highly sensitive on implicit sample selection effects made
in such an analysis. This paper attempts to resolve some of these
issues, and seeks to answer the question if AGN host galaxies inhabit
the same environs as their non-active counterparts with similar
properties. In contrast with existing studies however, to minimize
implicit selection biases, we choose two AGN samples each selected
from the two extremes of the AGN colour distribution that we match
with non-active comparison samples that have the same colour
distribution. Our method will increase the sensitivity of any
environmental signal that may otherwise be diluted by
noise induced by the averaging across a large baseline of galaxy type,
a problem which we consider implicit in some of the aforementioned
studies.

The layout of the paper is as follows. Section 2 describes the data
and sample selection criteria. Section 3 contains the large scale
clustering analysis for our active and non-active galaxy
environments. In Section 4 we analyse the mean density profiles around
galaxy groups. An analysis of the morphology of AGN hosts is described
in Section 5. In Section 6 we determine and compare properties of
galaxies in neighbourhoods of both non-active and AGN hosts. We
summarize the main conclusions of our study in Section 7.

The cosmological parameters adopted throughout this paper are: $\Omega = 0.3$,
$\Omega_{\lambda} = 0.7$  and $H_0 = 100 \kms Mpc^{-1}$.

\section{Data}

The Sloan Digital Sky Survey \citep[SDSS,][]{york00} has
mapped one-quarter of the entire sky in 5 optical bands and performed a deep 
redshift
survey of galaxies, quasars and stars. The fourth data release, DR4,
from SDSS provides a database of some $\sim$$500000$ galaxies with measured
spectra. The five filters \textit{u, g, r, i,} and \textit{z} cover
the entire wavelength range of the CCD response function \citep{fukugita96}.
The main galaxy sample is essentially a magnitude limited
spectroscopic sample (with a \cite{petro76}
magnitude) \textit{$r_{lim}$}$ < 17.77$, with most galaxies spanning a
redshift range $0 < z < 0.25$ with a median redshift of
0.1 \citep{strauss02}.

Several important properties have been derived for various
sub-samples of SDSS galaxies: gas-phase metallicities, stellar masses, 
indicators of recent major
star-bursts, current total and specific star-formation
rates, emission-line fluxes etc  \citep{brinch04,tremonti04} by spectroscopic 
targeting of the main galaxies as well as candidate regional members. A 
comprehensive study of SDSS type II AGN including classification scheme is 
undertaken in  \citet{kauff03a} based on the standard emission line ratio 
diagnostic diagrams.

\subsection{Sample Selection}

AGN are often separated into two categories: Type 1 AGN, where the black hole
and the associated stellar continuum and emission-line regions (both broad and
narrow) are viewed, and Type 2 AGN for which the broad emission line
region can not be viewed because of relative orientation of an
obscuring medium. In this paper we investigate the properties of AGN 
hosts by comparing them to matched
non-active galaxies. We present evidence of the nature of the AGN
trigger/fuelling mechanism by focusing on environment. Since the
observed properties of the host galaxies of type 1 AGN are often
contaminated by the contribution of the central nucleus, only type 2
AGN are included in this investigation, although we expect our results
to also apply to type 1 AGN environments because essentially they are
the same phenomenon.  The \cite{BPT81} line-ratio diagram allows for
the separation of type 2 AGN from normal star forming galaxies by
considering the intensity ratios of two pairs of relatively strong
emission lines.  The sample of Type 2 AGN of \cite{kauff03a} was
selected taking into account the relation between spectral lines,
$\rm [OIII]\lambda 5007$, $\rm H\beta$, $\rm [NII]\lambda 6583$ and $\rm H\alpha$
luminosities where an AGN is defined by

\begin{equation}
\log_{10}([OIII]/\rm H\beta) > 0.61/(\log_{10}(\rm [NII/H\alpha])-0.05)+1.3
\end{equation}

We use the type 2 AGN sample of \cite{kauff03a} as
well as a control sample from the main spectroscopic SDSS galaxy sample.

The colours of galaxies can be used to discriminate the stellar
populations and their evolution.  In galaxy clusters the larger
fraction of red galaxies indicate an old stellar population with
galaxies typically having a lower star formation rate (SFR), while
galaxies in poor groups and the field are bluer and have higher
SFR. Thus, from the type 2 AGN sample two sub-samples were selected at
the extremes of the colour distribution; a red sample targeting
potential cluster AGN and a blue sample, chosen to represent field
AGN.

\begin{figure*}
\includegraphics[width=180mm]{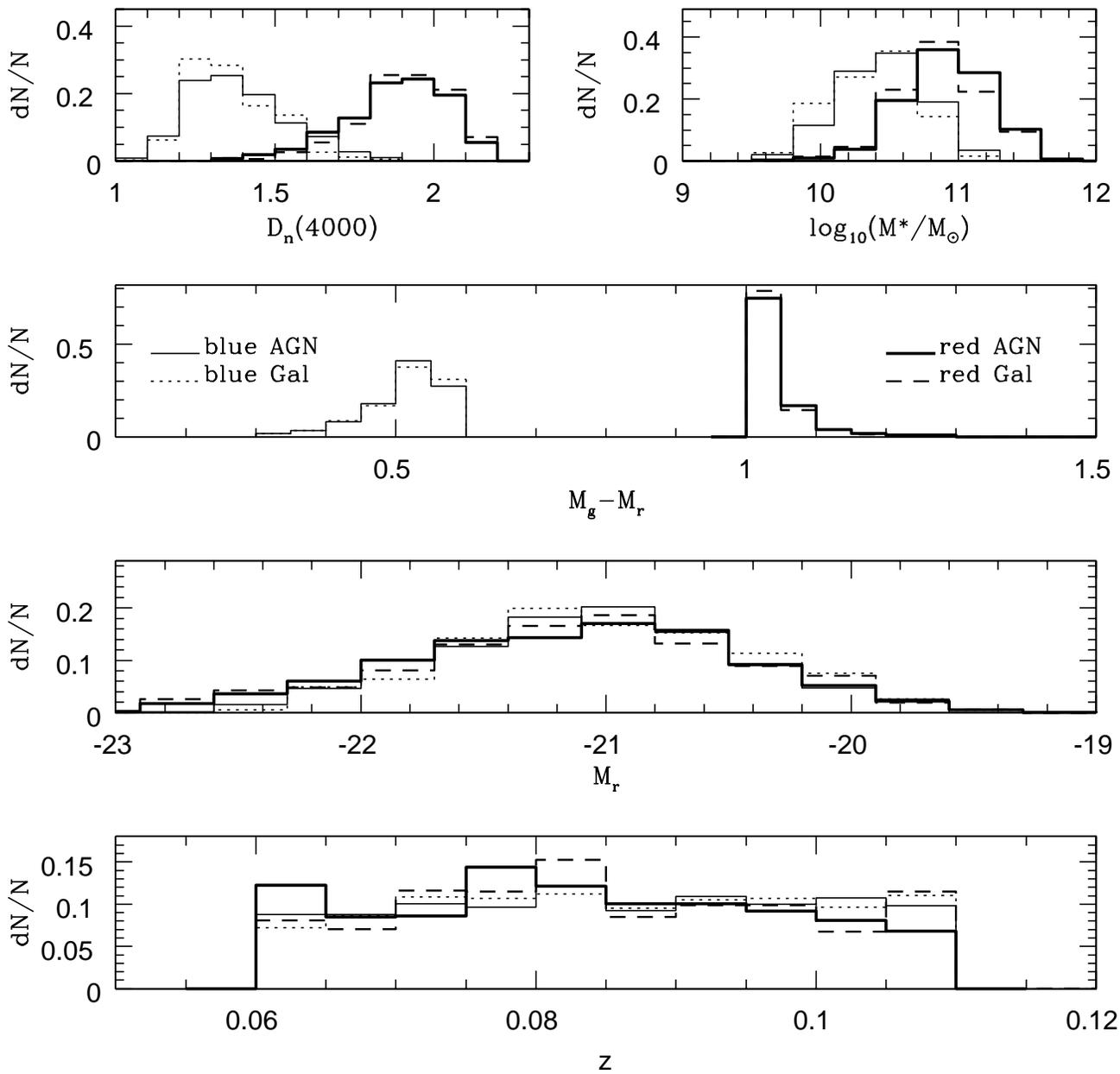}
\caption{Normalized distributions of matched system properties: red AGN (thick 
solid 
line) \& galaxies (dashed line) and blue AGN (thin solid line) \& galaxies 
(dotted 
line)}
\label{fig1}
\end{figure*}

We divide the AGN sample according to $\rm M_g-M_r$ colours using K and
extinction corrected absolute magnitudes \citep{blanton03}. Here the
red AGN are $\rm M_g-M_r > 1$ and the blue AGN are $\rm M_g-M_r < 0.6$. With
these colour cuts, we are considering approximately only $5\%$ of both
extremes of the galaxy colour distribution, within $0.06 < z <
0.11$. Each represent two very distinct populations of AGN host
galaxies.  The AGN samples are also restricted to have identical
redshift distributions.

We compare the results obtained for the active host galaxy environment
with two carefully selected control samples of non-active
galaxy environments. Each of the galaxy control samples is selected with 
properties
to match the AGN samples in redshift, luminosity, colour, stellar
mass and mean stellar age, measured by the 4000 {\AA} break strength,
distributions. These restrictions ensure that any difference in
environmental properties can only be attributed to nuclear
activity. We note that the stellar mass from SDSS AGN and galaxies
have been previously determined by \cite{kauff03b} using a method that relies 
on spectral
indicators of the stellar age and the fraction of stars formed in
recent bursts.  The break index $\rm D_n(4000)$ \citep{kauff02} is defined
as the ratio of the average flux density in the narrow continuum
bands (3850-3950 {\AA} and 4000-4100 {\AA}) and is suitably correlated
to the mean age of the stellar population in a galaxy and can be used
to estimate the star formation rate per unit stellar mass, $SFR/M*$,
\citep{brinch04}. The majority of star formation takes place
preferentially in galaxies with low $\rm D_n(4000)$ values.
Considering all these restrictions we have 1187 AGN and 1062 galaxies
for the red sub-samples and 1193 AGN and 1152 galaxies for the blue 
sub-sample (see Figure~\ref{fig1} to test the environmental dependence on the 
distribution of properties).

\section{Cross-correlation analysis}

Several studies indicate that on average, AGN and quasars populate
regions with density enhancement similar to those of typical
galaxies \citep{coldwell06}. The projected cross-correlation function
calculated for a sample of AGN with a control sample of similar
characteristics by \cite{Li06} has revealed that AGN and galaxies
inhabit dark matter halos of similar mass.  In this section we
explore the clustering properties between galaxies and blue and red
AGN hosts and compare these results with control samples.

The large-scale clustering of galaxies is very well characterized by
the two-point cross-correlation function, $\xi$.  We have calculated
the real-space, $\xi(r)$, cross-correlation function straightforward
using the two-dimensional standard estimator
$\Xi(\sigma,\pi)$ \citep{DP83}, where $\sigma$ is the perpendicular
distance and $\pi$ is the distance parallel to the line of sight.

\begin{equation}
\xi(\sigma,\pi)=\frac{D_iD_j(\sigma,\pi)N_{R_i}N_{R_j}}{R_iR_j(\sigma,\pi)N_{D
_i}N_{D_j}}-1, 
\end{equation}

where $D_iD_j$ and $R_iR_j$ are the measured number of data-data and
random-random pair counts respectively, binned as a function of the
separation of the two variables $\sigma$ and $\pi$, and $N_{D_i}$,
$N_{R_i}$ are the mean number densities of galaxies in the data and
random samples. The random samples are created to have 10 times more
objects, set to have the same radial and angular selection function as
do the real catalogues.  Integrating along the line of sight, where
$\pi_{max}= 60 \mpc$ \\

\begin{equation}
\Xi(\sigma)= 2 \int_0^{\pi_{max}} \xi(\sigma,\pi)d\pi ,
\end{equation}

we can estimate $\xi(r)$ by directly inverting $\Xi(\sigma)$ assuming
a step function $\Xi(\sigma)=\Xi_i$ in bins centred in $\sigma_i$ and
interpolating between values for $r= \sigma_i$ \cite{saunders92}.

\begin{equation}
\xi(\sigma_i)=
-\frac{1}{\pi}\sum_{j\ge i} 
\frac{\Xi_{j+1}-\Xi_j}{\sigma_{j+1}-\sigma_j}\ln( 
\frac{\sigma_{j+1}+\sqrt{\sigma_{j+1}^2-\sigma_i^2}} 
{\sigma_{j}+\sqrt{\sigma_{j}^2-\sigma_i^2}} )
\end{equation}

\begin{figure}
\includegraphics[width=90mm,height=90mm]{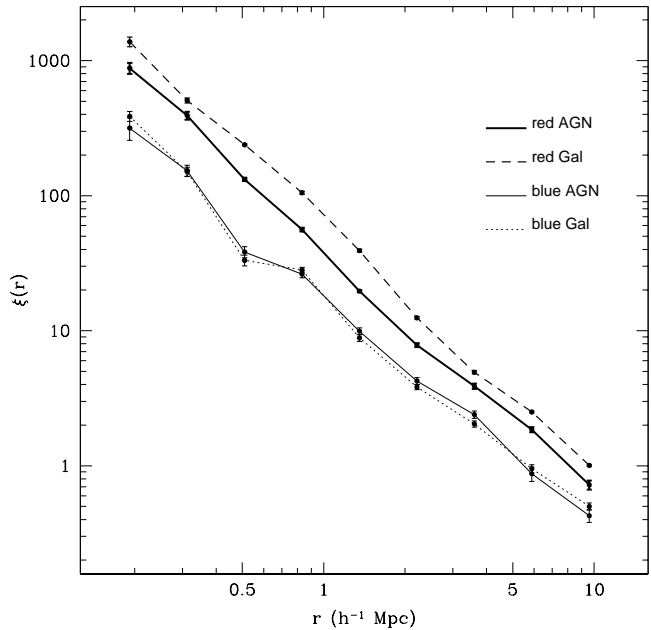}
\caption{Real-space cross-correlation functions: red AGN (thick solid line) 
\& galaxies (dashed line) and blue AGN (thin solid line) \& galaxies (dotted 
line)}
\label{fig2}
\end{figure}

The results are shown in Figure~\ref{fig2}. The error-bars were
calculated with the jackknife technique \citep{lupton93} by dividing
the sample into 10 sky regions of approximately similar area. It is not
unexpected that the red samples show stronger clustering than do the
blue samples. The red early type galaxies are known to occupy regions
of higher densities such as clusters or galaxy groups whereas the
blue galaxies tend to be isolated or in the outer regions of such
systems. A clear difference can be noticed between the red populations
of active and non-active galaxies: red AGN are found in environments
with lower densities at all scales than are their non-active counterparts. 
Such effect cannot be observed for the blue AGN sample which shows almost 
identical degree of large scale clustering as the galaxy comparison sample.
To further test the origin of the difference between active
and non-active red galaxies, in the following section the
density of AGN around groupings of galaxies is analysed.

Moreover, it is possible to observe that $\xi(r)$ shows two regimes
with a transition from the linear to the non-linear case. The shape of
$\xi(r)$ and specifically this transition could be related to the
nature of the Halo Model \citep{CS02} whereby in the inner regions of
the correlation function galaxies belong to a same halo while for the
outer regions the galaxies may populate different halos. The
transition between both regimes appears to be sensitive on the colour
and luminosity of the galaxies.  For the blue systems the transition
is seen to occur at smaller scales than for red ones, because red
galaxies are generally found within galaxy systems, while blue
galaxies are more often found in isolation.
 
\section{Density of AGN around galaxy groupings}

To determine the density of AGN within high density environs we use
galaxy groups taken from \cite{berlind08}. These galaxy groups
were identified from the SDSS using an optimized redshift-space
friends-of-friends algorithm (FOF). The FOF linking lengths were
selected to group together galaxies that occupy the same dark matter
halos (for details see \cite{berlind06}).

\begin{figure}
\includegraphics[width=85mm,height=95mm]{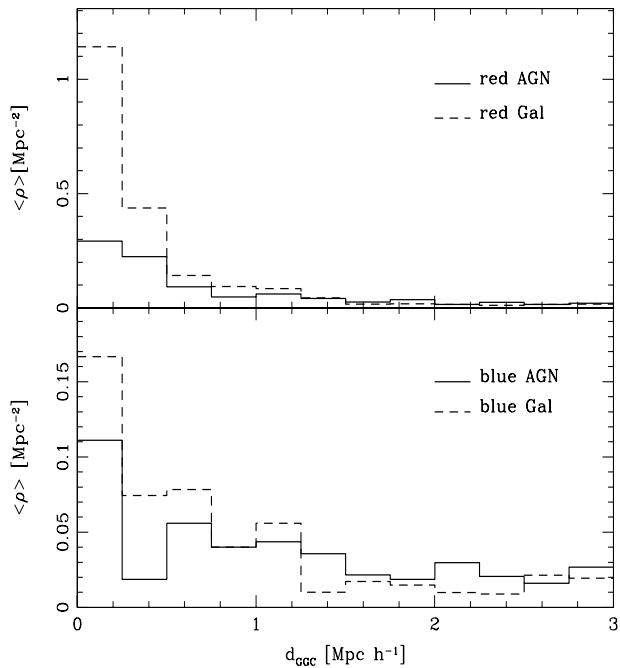}
\caption{Mean density of galaxies (dashed lines) and AGN
  (solid lines) around Berlind's galaxy groups with more than 15
  members. The red samples are in the upper panel and the blue ones
  in the bottom panel}
\label{fig3}
\end{figure}

In this section we calculate the normalized mean density profile of
AGN and non-active galaxies within a projected distance, $d_{GGC} <
3 \mpc$ and with radial velocity differences, $\Delta V < 2000 \kms$
with respect to the galaxy group centres. We conducted tests dividing 
the galaxy groups
according to their richness expressed as the number of member
galaxies. Figure~\ref{fig3} shows the mean density profiles around
galaxy groups with more than 15 members. The expected high density of
red galaxies around group centres is noticeable, while for red
AGN, it is approximately four times lower within $250 \kpc$. The excess of red
galaxies relative to red AGN around galaxy groups extends up to
$\sim 1 \mpc$. The mean density of the blue AGN on the other-hand is
considerably lower and only a small difference between blue galaxies
and blue AGN can be observed.

This result indicates that even though red AGN are associated with
much richer environments than are blue AGN, both samples are
distributed differently compared to non-active red galaxies. That is to say,
even when morphological segregation \citep{Dress80} is taken into
account, red AGN are still preferentially located away from the
centres of galaxy groups, a tendency that increases with group
richness. More evidence of the nature of this effect is presented in
the next section where we analyze the distribution of the
morphological types of our galaxies.

\begin{figure}
\includegraphics[width=85mm,height=95mm]{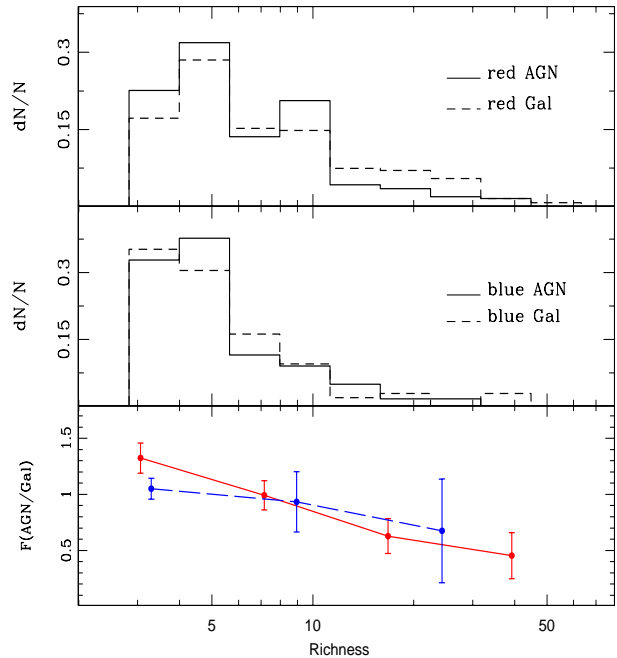}
\caption{Relative distribution of AGN (solid lines) and galaxies (dashed 
lines) 
as a function of richness (in logarithmic scale), within $\rm d_{GGC} < 0.75 \mpc$ 
and
$\Delta V < 2000 \kms$ from the galaxy group centre. The red samples are in 
the upper panel and the blue ones in the middle panel. 
Bottom panel: Fraction of AGN with respect to the galaxies, $\rm F(AGN/Gal)$, 
for red (red solid line) and blue (blue long-dashed line) samples respectively. }
\label{fig3b}
\end{figure}

Additionally, it is important to emphasize that the fraction of AGN decreases
with number of galaxy group members and this effect is more evident for the 
red AGN. To illustrate this behaviour we select all the galaxies and AGN 
within a group centre projected distance $\rm d_{GGC} < 0.75 \mpc$ and
$\Delta V < 2000 \kms$ where the lack of AGN is more pronounced as can be 
seen
in Figure~\ref{fig3}. 
We show the relative distribution of AGN and galaxies as a function of 
richness,
given by the number of galaxy group member, in Figure~\ref{fig3b} for red 
samples 
(top panel) and blue samples (middle panel). The diminishing number of AGN 
for
galaxy systems with more than 10 members is perceptible for red AGN while for
blue AGN this effects does not seem to be obvious.

In order to quantify the dependence of the AGN presence with the
richness of galaxy groups we calculate the fraction of AGN with
respect to the galaxies, $\rm F(AGN/Gal)$, for red and blue samples
respectively as it is shown in Figure~\ref{fig3b} (bottom panel).  It
is possible to observe a clear deficiency of red AGN in the richest
galaxy groups with respect to the red non-active galaxies. There is a
strong relation between red AGN occurrence and the number of galaxy
group members, while for blue AGN the relation is weak. Moreover, the
evidence, that the fraction of blue AGN remains almost constant with
the richness is a signal that dominates the errors induced by the
smaller number of blue AGN and galaxies in our rich galaxy groups.

Our result for red AGN are consistent with those of \cite{PB06}. They
found an anti-correlation between the total fraction of AGN, with respect to 
the
cluster members, and the velocity dispersion for two cluster samples.
Their conclusion suggest that the increasing fraction of AGN vs. the 
decreasing
velocity dispersion is related to the merger rate since the merger rate has an 
inverse cubic dependence on the velocity dispersion of the cluster or group, 
so it could explain the lack of the AGN fraction in richer systems. 

\section{Morphologies of AGN hosts}

In contrast to bright Type I AGN, the morphologies of the host galaxies of 
Type II AGN are usually easily observable. The morphologies can be quantified 
using S\'{e}rsic profile \citep{sersic63} that describes how the luminosity 
of a galaxy varies with distance from its centre, a generalisation of 
the de Vaucouleurs and Freeman laws. 

The S\'{e}rsic profile has the form

\begin{equation}
\ln I(R)= \ln I_{0} -kR^{1/n},
\end{equation}
where $I_{0}$ is the intensity at $R = 0$. The parameter $n$, called
the S\'{e}rsic index, controls the degree of curvature of the
profile. The best-fit value of $n$ correlates with galaxy size and
luminosity, such that bigger and brighter galaxies tend to be fit with
larger values of $n$ \citep{caon93, YC94}. Most galaxies are fit by
S\'{e}rsic profiles with indices in the range $0.5 < n < 10$. Setting
$n = 4$ gives the de Vaucouleurs profile which is a good description
of giant elliptical galaxies. Setting $n = 1$ gives the Freeman
exponential profile which is a good description of the light
distribution of both disk galaxies and dwarf elliptical galaxies.

\begin{figure}
\includegraphics[width=85mm,height=95mm]{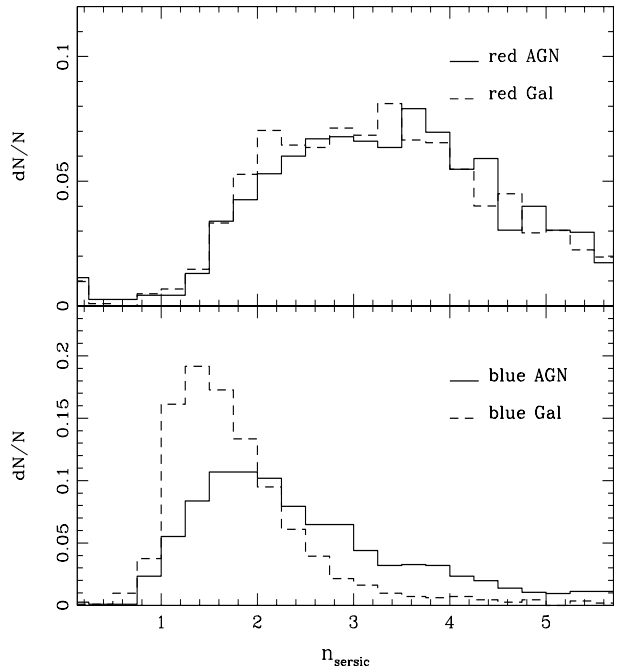}
\caption{The distribution of the S\'{e}rsic index for red (top) and blue 
(bottom) samples for galaxies (dashed) and AGN (solid) lines. The
S\'{e}rsic index provides a measure of the morphological type of a galaxy
with lower numbers corresponding to spirals and higher numbers corresponding 
to
ellipticals.}
\label{fig4}
\end{figure}

Figure~\ref{fig4} illustrates the distribution of the S\'{e}rsic
indices for red and blue AGN hosts relative to their non-active
comparison samples. It is clear that red AGN have essentially the same
morphological make-up as do the red non-active sample
providing the final linchpin that the relative absence of AGN in cores of
clusters and groups \emph{cannot be explained simply as a consequence of
morphological segregation}. Somewhat surprisingly, the morphologies of
blue AGN are considerably different from the blue non-active
galaxies despite their environs being almost identical at least in
terms of their density. This finding leads to the question of whether
or not the properties of the galaxies in these neighbourhoods are also
different, a question we address in the following section.

However, we note from the distribution of the S\'{e}rsic indices
of both samples that it is apparent that nuclear-activity is a
rare event in spiral disks and dwarf ellipticals. On the other hand, extended 
blue
ellipticals seem to promote nuclear-activity. Blue
early-type galaxies have been shown to inhabit low density
environments away from cluster centres \citep{bamford09} just like
blue spirals, suggesting that the morphology of the host does play
a role in triggering or sustaining nuclear activity.

\section{Properties of the galaxy neighbourhoods}

\subsection{Dependence on host properties}

The method commonly adopted for deriving a galaxy's star formation rate is 
based on the
modelled contribution of both the nebular emission by HII regions and
diffuse ionized gas that are often combined and described in terms of an 
effective
metallicity, ionization parameter, dust attenuation parameter at 5500 {\AA}, 
and
dust to metal quotient \citep{BC93,charlot02}.  Taking into account these
factors, \citep{brinch04} provide accurate estimates of total star
formation rates that are free from aperture bias. 

Considering the results of the previous sections, one might expect
that similar trends will be reflected in the properties of the
neighbouring galaxies. Following the analysis of \cite{coldwell06} we
analyze the colours $\rm M_u-M_r$ and the logarithmic specific star
formation rate $\rm \log_{10} (SFR/M^*)$ [$\rm \log_{10}$ yr$^{-1}$], 
where $\rm M*$ is the
estimated stellar mass, for galaxies surrounding the different targets.

\begin{figure}
\includegraphics[width=85mm,height=90mm]{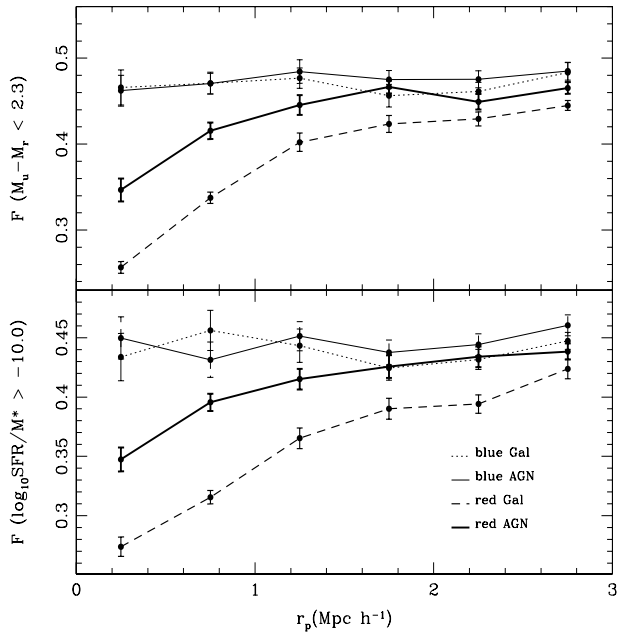}
\caption{Fraction of $\rm M_u-M_r$ blue (top panel) and star-forming (bottom 
panel) systems 
 vs. $r_p$: red AGN (thick solid line) \& galaxies (dashed line) and blue AGN 
(thin solid line) \& galaxies (dotted line) }
\label{fig5}
\end{figure}

In order to quantify any excess of blue (star forming)
galaxies in the AGN environs, we calculate the fraction of
galaxies bluer than $\rm M_u-M_r < 2.3$ approximately corresponding to
the mean $\rm M_u-M_r$ for the spectroscopic survey at $z < 0.1$ as well as the
fraction of star forming galaxies given by $\rm \log_{10}(SFR/M^*) > -10.0$, both 
shown 
in Figure~\ref{fig5}.
Tracer galaxies are selected to have projected distances $r_p < 3 \mpc$
and radial velocity differences $\Delta V < 1000 \kms$, relative to the
target system. 

\begin{figure}
\includegraphics[width=85mm,height=90mm]{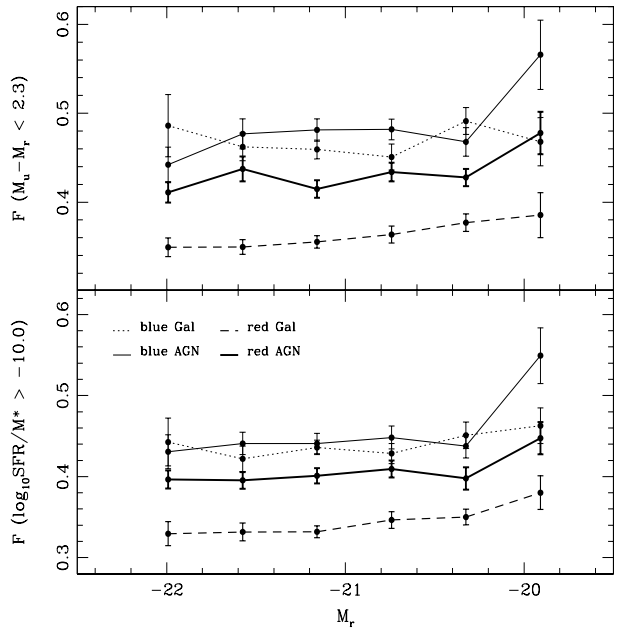}
\caption{Fraction of blue (top panel) and star-forming (bottom panel) systems 
vs. $\rm M_r$: red AGN (thick solid line) \& galaxies (dashed line) and blue AGN 
(thin solid line) \& galaxies (dotted line)}
\label{fig6}
\end{figure}

Figure ~\ref{fig5} shows that the environs of both active and
non-active blue systems are practically indistinguishable, whereas,
for the red systems significant differences are apparent. The environs
of red AGN are populated by bluer galaxies with stronger star
formation compared to the red non-active galaxies. The difference is
approximately 3 $\sigma$ within $r_p < 1 \mpc$ and this signal remains
strong beyond $r_p = 2 \mpc$.  This result is consistent
with \cite{coldwell06} who find a higher fraction of blue star forming
galaxies associated with AGN environs without such a host colour
discrimination. We observe no similarity in the excess of starforming
galaxies around the sample of red AGN compared to the non-active red
galaxies (see Figure~\ref{fig6}).

We also calculate the fraction of blue and star forming galaxies as a
function of the luminosity, $\rm M_r$, of the targets
within $r_p < 2 \mpc$ and $\Delta V < 1000 \kms$ from the targets. 
We observe an excess of star-forming
galaxies around red AGN compared with their non-active counterparts while the
environment of blue systems remain clearly similar. We also notice a weak
relation between the target luminosity and the calculated fraction. In
general, low luminosity AGN and non active galaxies tend to populate
regions where larger numbers of blue star forming galaxies are also
present.

\begin{figure}
\includegraphics[width=85mm,height=90mm]{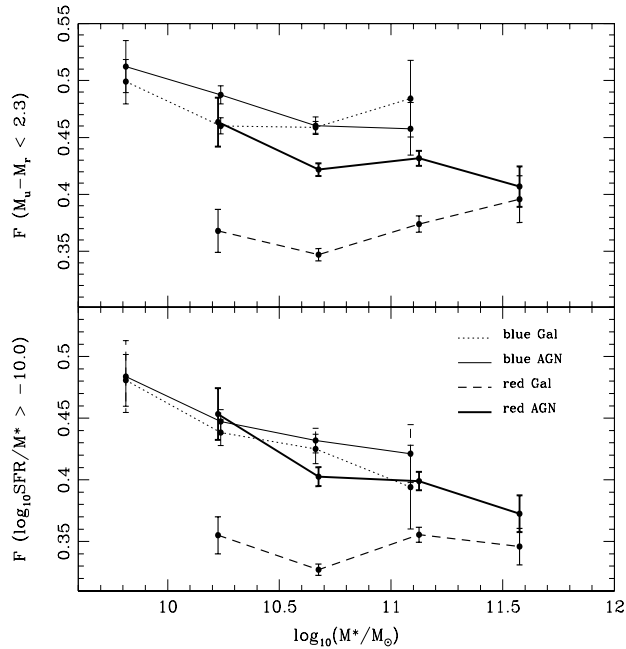}
\caption{Fraction of blue and star forming systems as a function of the 
stellar mass: red AGN (thick solid line) \& galaxies (dashed line) and blue 
AGN (thin solid line) \& galaxies (dotted line)}
\label{fig7}
\end{figure}

Moreover, the mass-luminosity relation in galaxies is well known. However this
relation has a significant spread that depends on the method used to
derive the two parameters. Taking into account this caveat, at our
level of completeness, we calculated the fraction of blue and star
forming galaxies as a function of the stellar mass, 
$\rm \log_{10}(M^*/M_{\odot})$,
of the targets, within $r_p < 2 \mpc$ and $\Delta V < 1000 \kms$. 
The results appear in Figure~\ref{fig7} where a significant difference is 
observed
between the environments of the red systems as function of their host
stellar mass. This difference is larger for the less massive red AGN
which show environmental properties not too dissimilar to those of
more massive blue systems. One can see from Figure~\ref{fig7} that the
fraction of galaxies with $\rm M_u-M_r < 2.3$ and $\rm \log_{10}(SFR/M^*) >
-10.0$ around blue systems and red AGN decrease with $\rm M^*$. No such
trend is visible for the non-active red galaxies.
 
In addition, the values which characterize the neighbours of red AGN are much 
more comparable to those for the blue AGN and blue non-active galaxies. 
Even though they are not identical, the environments of the blue systems, 
AGN and 
non-active galaxies are practically indistinguishable within our errors.

\subsection{Dependence on AGN activity}

The relation between the AGN activity and host properties is a
particularly interesting connection for study. To date the most useful
AGN activity parameter is the luminosity of the $\rm [OIII]\lambda 5007$
line ($\rm L[O III]$ ), calculated by \cite{kauff03a}. Although this line
can be excited by massive stars as well as by AGN, it is known to be
relatively weak in metal-rich, star-forming galaxies. The $\rm [O III]$
line also has the advantage of being strong and easily detected in
most galaxies, however, the narrow-line emission is likely to be
affected by dust within the host galaxy \citep{kauff03a} and thus it
is important to correct for the effects of extinction.  For AGN in
SDSS, \cite{kauff03a} estimated the extinction using the Balmer
decrement, finding that the best approximation for a dust correction
to $\rm L[O III]$ is based on the ratio $\rm H\alpha /H\beta$.

\begin{figure*}
\includegraphics[width=175mm]{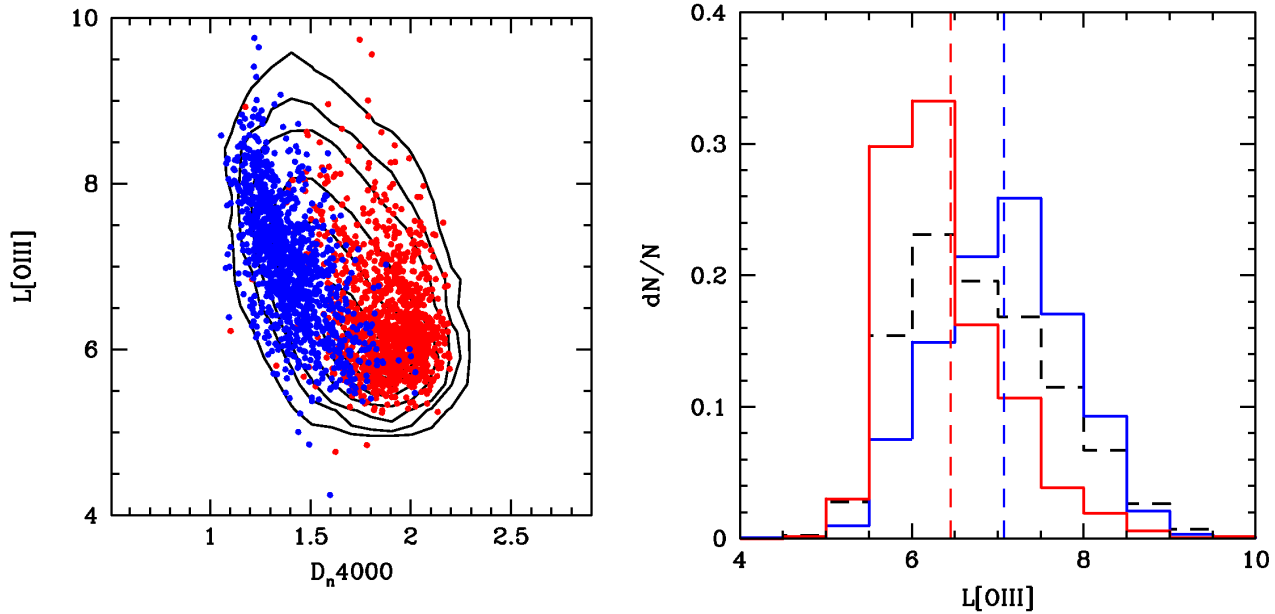}
\caption{
{\bf Left}:$\rm L[O III]$ vs. $\rm D_n4000$ for the total samples of AGN (contours), blue 
and red 
AGN (dots).  {\bf Right}: $\rm L[O III]$ distributions for the total sample of 
AGN
(black dashed line), red AGN (solid red line) and blue AGN (solid blue line).
The red and blue vertical dashed lines correspond to the mean values of $\rm 
L[O III]$
of the red and blue AGN samples, respectively.}
\label{fig8}
\end{figure*}

By inspection of the relation between different parameters measured
for the AGN hosts and the $\rm L[O III]$ luminosities we find a linear
correlation between $D_n(4000)$ and $\rm L[O III]$. Although we can not rule 
out some contamination to the $\rm L[OIII]$ values from star forming regions, this result indicates
that AGN with higher values of $\rm L[O III]$ most likely contain
younger stellar populations than those with lower values of $\rm L[O III]$
which are in a more evolved stage whereby gas deficiency causes a marked diminution 
in nuclear activity.  Furthermore a separation by
galaxy colour implies a smooth separation by $\rm L[O III]$ as it is shown
in Figure~\ref{fig8} (left panel). Thus a wide
separation of two populations is visible, where the redder AGN hosts have very 
low
values of $\rm L[O III]$ compared with the bluer AGN.

\begin{figure}
\includegraphics[width=90mm,height=90mm]{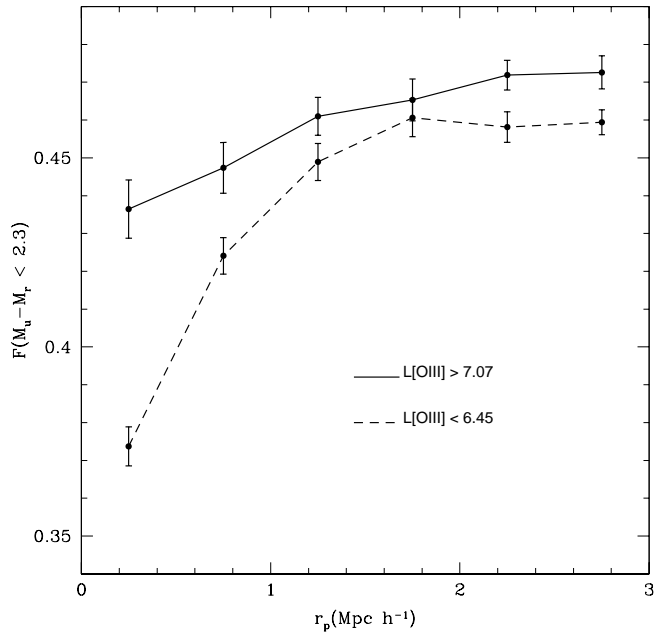}
\caption{Blue galaxy fraction as a function of projected distance
to weak (dashed line) and powerful (solid line) AGN. The values
separating the powerful and weak AGN are taken from the mean values
corresponding to the $\rm L[OIII]$ distribution of blue and red AGN
respectively (see Figure~\ref{fig8}) }
\label{fig9}
\end{figure}

Taking into account the correlation between the $\rm M_g-M_r$ colours and
$\rm L[O III]$ showed in Figure~\ref{fig8} (left panel) we complete the 
analysis and select AGN samples with
extreme values of $\rm L[O III]$. We consider weak AGN to be those
with $\rm L[O III] < 6.45 L_\odot$ and powerful AGN to be those with
$\rm L[O III] > 7.07 L_\odot$ which are the mean of the $\rm L[O III]$
values for the red and blue AGN samples, respectively (see
Figure~\ref{fig8}, right panel). Again, we calculate the fraction of
blue galaxies within $r_p < 3 \mpc$ and $\Delta V < 1000 \kms$ of the
host system. The results are showed in Figure~\ref{fig9}. As expected
from the previous results, the powerful AGN are located in regions
over populated by blue star forming galaxies with respect to the weak
AGN. Therefore, our results seem to indicate that the power of the AGN
activity is strongly dependent on the environment of the host
galaxies.

\section{Summary}

In this paper we investigate the environs that host Type II AGN. In contrast 
with existing studies we select two samples at the extremes of the colour 
distribution that we match to comparison samples of non-active galaxies. 
By focusing on the extreme color ends of the AGN population, we are able to 
avoid averaging results 
across long baselines and so too, associated biases implicit in some of the 
earlier studies.

We find that the density distribution expressed by the
cross-correlation function is almost identical for blue, active and
non-active host galaxies. In contrast, red active galaxies inhabit
environments less dense compared to their red non-active counterparts. In 
spite
of this, their environs are still considerably denser than those of
blue hosts. In addition, two regimes are observed in the shape of the real-space
cross-correlation function, $\xi(r)$, with a transition between its inner
and outer regions related with the property that galaxies belong to only one 
or several halos. This turning point seems to occur at smaller scales for
blue systems, which are more often found in isolation, with respect to the 
red ones generally found within galaxy groups or clusters.  

Moreover, the distribution of red AGN relative to galaxy groups and
clusters shows a significant paucity or underpopulation at their
centres. This cannot be explained by morphological segregation, since
the morphological distributions of active and non-active red systems
are identical. Additionally, we notice than the fraction of AGN 
decreases with number of galaxy group members and this effect is more
evident for the red AGN. We find a strong relation between red AGN
occurrence and the number of galaxy group members, while for blue AGN
the relation is weak. This could suggest that the fraction of blue AGN
remain almost constant with the richness taking into account the
errors of the fraction due the small number of blue AGN and galaxy in
rich galaxy groups.

On the other hand the active and non-active blue systems have identical 
environments but markedly different morphological distributions.
We note from the distribution of the S\'{e}rsic indices of the samples 
that it is apparent that nuclear-activity is a rare event in spiral disks 
and dwarf ellipticals. 
The population of blue AGN hosts has a considerable contribution of ellipticals 
which are thought to be late stage mergers with already established morphology 
but remaining photometric anomalies \citep{MAF04}.
Blue early-type galaxies have been shown to inhabit low density
environments away from cluster centres \citep{bamford09} just like
blue spirals, suggesting that the morphology of the host does play
a role in triggering or sustaining nuclear activity.

These results indicate that blue and red AGN might have intrinsically
different formation mechanisms, suggestive of a dichotomy in the AGN
population. Obviously, one could argue that such a dichotomy in the 
environment is a natural
consequence of the sample selection derived from the anti-poles of the
AGN population.  For example, according to a review
by \cite{storchi08}, nuclear activity outlives star formation and in
some cases its onset might be even delayed until after the star
formation period is triggered in the same episode. In such a scenario
the blue AGN in our sample would mark mainly the onset of this
activity and red AGN its final stage. In this sense one could
interpret our blue sample to indeed fulfill this general idea: (i)
morphology plus colour combination suggesting recent galaxy
interactions; (ii) high star formation rate; (iii) young stellar
population; (iv) strong nuclear activity; and (v) scaling of nuclear
power with age of the stellar population indicating gradual depletion
of fuel.

The $\rm L[OIII]$ vs $\rm D_n(4000)$ distribution (nuclear power vs age of stellar 
population) of all AGN appear to naturally follow the trend set by our blue 
sample with exception of the extremely red AGN. The red AGN in our sample are 
much older than would otherwise be suggested by an extrapolation of the trend 
marked by the blue and intermediate colour AGN. Thus, to belong to the same 
population as the blue AGN their stellar population would need to age at an 
accelerated rate. We revisited some possible scenarios.

In a first scenario, a blue galaxy with an AGN would enter a cluster, 
be stripped of its gas envelope, leading to starvation \citep{BNM00}, while the 
gas content at its center is withheld at the deepest position in the galaxy 
potential well.
Thus, the AGN could continue to be fed while the star-formation is effectively 
halted throughout most of the galaxy. The problem is that the stripping of gas 
doesn't accelerate aging, but simply truncates the starformation. It means that 
galaxies wouldn't show ongoing star-formation but would still have relatively 
young populations considering the short duration of a typical burst of nuclear 
activity.

In the second scenario, the extremely red AGN have a separate 
trigger mechanism that occurs when an already dormant AGN enters into the strong 
gravitational potential of a galaxy cluster. The potential is known to strip 
some of the entering galaxies of their gas, however, not all of them. The 
observations of brightest 
cluster galaxies have found recent starformation in a third of those cluster 
core objects \citep{Lou08} suggesting that not all of the cluster galaxies are 
stripped of their gas when passing through the cluster potential. 
In this context, some galaxies will be stripped completely, some loose the gas 
in outer regions and some others may capture some of this deposited gas from 
the intracluster medium. Thus, an old, red, massive galaxy containing a 
dormant super massive black hole might acquire new gas supplies from the 
intracluster medium.  The gravitational potential of the cluster would assist 
in efficient transportation of the captured as well as galaxy's own gas 
component into the central black hole, triggering a burst of nuclear activity 
and resurrection of the AGN. In this scenario the red AGN correspond to a 
recycled species.

\section{Acknowledgments}

We thank the anonymous Referee for helpful suggestions and comments.
This research was partially supported by grants from CONICET, Agencia
C\'ordoba Ciencia and the Secretar\'{\i}a de Ciencia y T\'ecnica de la
Universidad Nacional de C\'ordoba. 

Funding for the SDSS and SDSS-II has been provided by the Alfred P. Sloan 
Foundation, the Participating Institutions, the National Science Foundation, 
the U.S. Department of Energy, the National Aeronautics and Space Administration, 
the Japanese Monbukagakusho, the Max Planck Society, and the Higher Education 
Funding Council for England. The SDSS Web Site is \emph{http://www.sdss.org/}.

The SDSS is managed by the Astrophysical Research Consortium for the Participating 
Institutions. The Participating Institutions are the American Museum of Natural 
History, Astrophysical Institute Potsdam, University of Basel, University of 
Cambridge, Case Western Reserve University, University of Chicago, Drexel University,
Fermilab, the Institute for Advanced Study, the Japan Participation Group, Johns 
Hopkins University, the Joint Institute for Nuclear Astrophysics, the Kavli Institute 
for Particle Astrophysics and Cosmology, the Korean Scientist Group, the Chinese 
Academy of Sciences (LAMOST), Los Alamos National Laboratory, the Max-Planck-Institute 
for Astronomy (MPIA), the Max-Planck-Institute for Astrophysics (MPA), New Mexico 
State University, Ohio State University, University of Pittsburgh, University of 
Portsmouth, Princeton University, the United States Naval Observatory, and the University 
of Washington.

{}

\label{lastpage}

\end{document}